# Latent mechanisms of polarization switching from *in situ* electron microscopy observations

Reinis Ignatans,[1] Maxim Ziatdinov,[2,3,a] Rama Vasudevan,[3] Mani Valleti,[4] Vasiliki Tileli,[1] and Sergei V. Kalinin,[3,b]

[1] Institute of Materials, École polytechnique fédérale de Lausanne, Station 12, 1015 Lausanne, Switzerland

[2] Computational Sciences and Engineering Division, Oak Ridge National Laboratory, Oak Ridge, TN 37831, USA

[3] Center for Nanophase Materials Sciences, Oak Ridge National Laboratory, Oak Ridge, TN 37831, USA

[4] Bredesen Center for Interdisciplinary Research, University of Tennessee, Knoxville, TN 37996, USA

*In situ* scanning transmission electron microscopy enables observation of the domain dynamics in ferroelectric materials as a function of externally applied bias and temperature. The resultant data sets contain a wealth of information on polarization switching and phase transition mechanisms. However, identification of these mechanisms from observational data sets has remained a problem due to a large variety of possible configurations, many of which are degenerate. Here, we introduce an approach based on a combination of deep learning-based semantic segmentation, rotationally invariant variational autoencoder (VAE), and non-negative matrix factorization to enable learning of a latent space representation of the data with multiple real-space rotationally equivalent variants mapped to the same latent space descriptors. By varying the size of training sub-images in the VAE, we tune the degree of complexity in the structural descriptors from simple domain wall detection to the identification of switching pathways. This yields a powerful tool for the exploration of the dynamic data in mesoscopic electron, scanning probe, optical, and chemical imaging. Moreover, this work adds to the growing body of knowledge of incorporating physical constraints into the machine and deep-learning methods to improve learned descriptors of physical phenomena.

[a] ziatdinovma@ornl.gov
[b] sergei2@ornl.gov

Recent advances in i*n situ* scanning transmission electron microscopy (STEM) allow for unprecedented observations of the dynamics of materials in response to external chemical, bias, and temperature stimuli with near atomic resolution. For instance, liquid phase and electrochemical STEM have enabled observation of battery devices operation,[1] morphogenesis of the electrodeposition and electroplating,[2] and structural evolution during electrocatalysis.[3] Environmental STEM has offered insights into the mechanisms of nanoparticle and nanowire growth.[4] Finally, elementary mechanisms of beam-induced reactions in layered dichalcogenides and graphene,[5-14] as well as certain bulk materials,[15-17] have been obtained at the atomic level.

These dynamic observations have also been observed in ferroelectric materials. These materials often elicit fascinating bias-induced behaviors and form multiple domain structures, with the complexity and morphology determined by intrinsic factors such as the number of possible domain variants and extrinsic factors such as impurities, external strain, and depolarizations fields.[18-20] Correspondingly, real-time observations provide insight into the physical mechanisms of nanoscale switching phenomena including domain nucleation and growth, domain wall motion, and domain interactions with interfaces and local defects.[21-27]

However, progress on understanding such complex dynamic phenomena in ferroelectrics has been limited by available techniques and the analysis tools employed. For instance, the vast majority of available observational data were mostly evaluated qualitative or simply analyzed via macroscopically averaged descriptors such as domain wall velocities or domain fractions. It would be pertinent to study individual moving interfaces to reveal the underlying mechanisms controlling such movement. The reason for not doing so is generally a lack of suitable analysis routines that can automate this process.

Here, we introduce an approach for analysis of the dynamics of bias- and temperature-induced transitions using a deep learning workflow based on semantic segmentation for identification of relevant structural features via rotationally invariant variational autoencoder. In this approach, one projects the high dimensional configurational space of domain morphologies on a low-dimensional continuous latent space allowing for possible orientation variants, thereby effectively encoding the possible domain configuration. The complexity of description can be tuned via the length-scale of sub-images, providing insight into the corresponding physical mechanisms. We further illustrate that the latent space of the variational autoencoder contains regions of both physical and unphysical domain structures. This, in turn, offers opportunities to explore evolution of the system via associated latent representations.

As a model system, we selected a proper ferroelectric material, $BaTiO_3$, whose domain evolution upon electric field biasing is well established. Positive and negative voltages were applied to a thin lamella sample using a triangular waveform in the polarization direction; the thin $BaTiO_3$ sample was mounted on a microelectromechanical systems (MEMS)-based device secured in a TEM holder.[28] The sample was heated to a temperature slightly under the Curie temperature where the 90º domains were stable and the response of the needle-shaped nanodomains to changes in the magnitude and direction of the induced electric field could be analyzed.[29]

Figure 1 shows a series of bright field (BF) STEM images that illustrate the evolution of the domain structure in BaTiO$_3$ during polarization switching at 130 ºC. The lines of dark contrast in the images correspond to domain walls; the random distribution of dark spots correspond to localized areas of built-up contamination on the BaTiO$_3$ sample. Two perpendicular 90º domains are observed at 0 V within the parent domain, Fig. 1(a). Upon positive biasing, Fig. 1(b)-(d), domains with a polarization vector aligned parallel to the direction of the applied electric field nucleate and grow while those with perpendicular polarization vectors are annihilated. In Fig, 1(e), the electric field is twice as strong as that in Fig. 1(d); however, the domain structure is nearly identical indicating complete switching. For an opposite electric field direction, Fig. 1(f), new needle-like domains with opposite polarization nucleate and the domain structure has reversed direction by 90°.

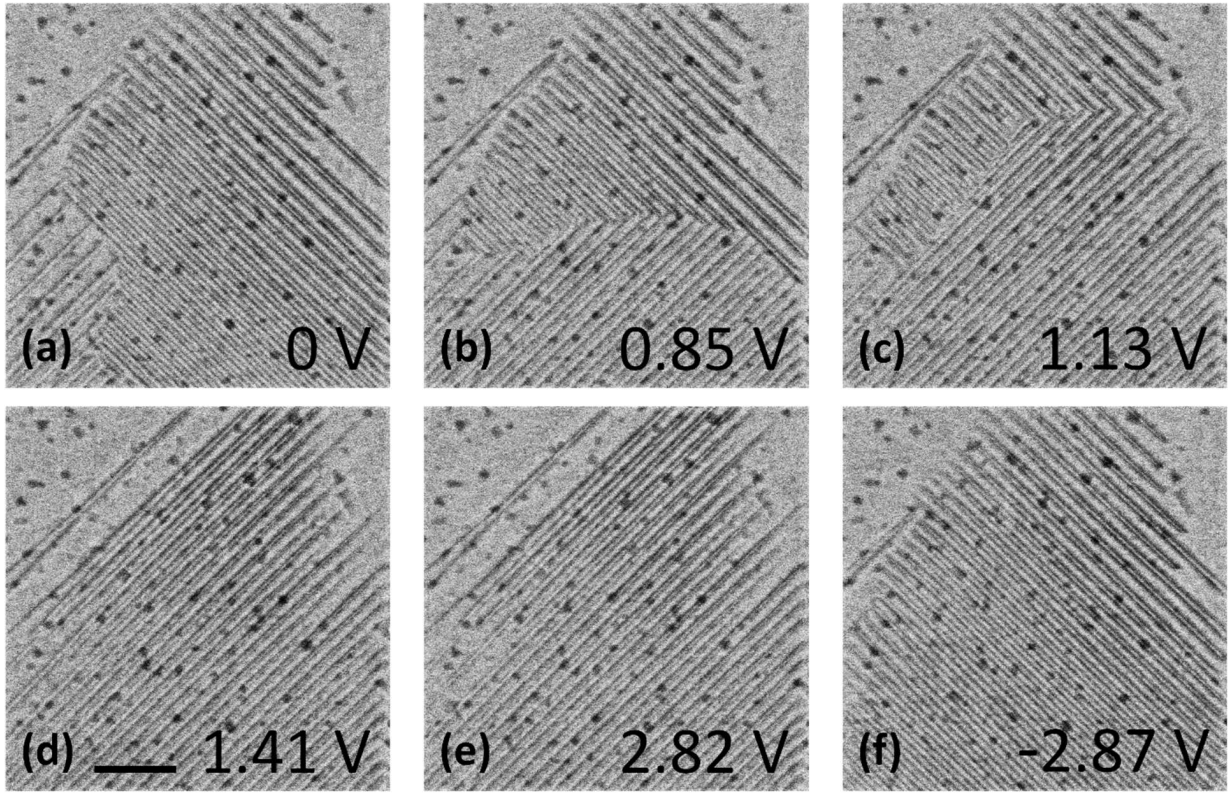

**Figure 1.** BF-STEM images showing domain evolution at different applied voltages. Scale bar (shown in (d)) is 500 nm.

Previously, we developed an approach for analyzing domain structure evolution during phase transformation using Gaussian mixture model (GMM) clustering of sub-images to identify possible classes of the domains, with subsequent dimensionality reduction to explore the dynamics of class population during an experiment.[30] These components in turn provide insight into the dynamics of the phase transition. However, the use of discrete classes limited this approach to a relatively small sub-image size since larger sub-images give rise to a rapidly growing number of discrete classes necessary for parameterization. An alternative approach is developed based on rotationally invariant variational autoencoders (rVAE), a class of unsupervised machine learning (ML) methods projecting discrete large-dimensional spaces on a continuous latent space. Previously, we applied the rVAE approach to explore the evolution of atomic-scale structures in graphene under electron beam (e-beam) irradiation[31] and analysis of domain wall dynamics in piezoresponse force microscopy[32]. Here, we demonstrate that this approach can be extended to explore domain evolution mechanisms via detailed analysis of the latent spaces of rVAE.

Prior to the rVAE analysis, we explored several strategies for pre-processing the experimental images to increase the visibility of the domain patterns. One approach was based on using classical image analysis tools such as resizing, blurring, and thresholding; however, these analyses are generally manpower intensive and not universal, leading to operator-dependent outcomes. Thus, to improve the identification of domain structures, we developed a deep convolutional neural network (DCNN)-based workflow for semantic segmentation of the images. In this approach, the human operator labels several of the images in terms of domain structures, backgrounds, and (optionally) contamination particles. The labeled data set is augmented and used to train the DCNN with the raw images as features and manually-defined classes (domain wall, background, contaminant) as labels. Once trained, the network can be used for analysis of the full image stack.

A stack of randomly cropped images was generated from several labeled image sequence frames and were used to train the U-net-like DCNN.[33] During training, the images were further augmented on-the-fly via random 90º rotations, horizontal/vertical flipping, zooming-in, and contrast changes to account for potential changes in the physical structure of domains (e.g., orientation and size) as well as potential changes in imaging conditions. The cross-entropy loss was optimized during DCNN training using the Adam technique [34] with a learning rate of 0.001. The home-built AtomAI software was used for both DCNN training and predictions.[35]

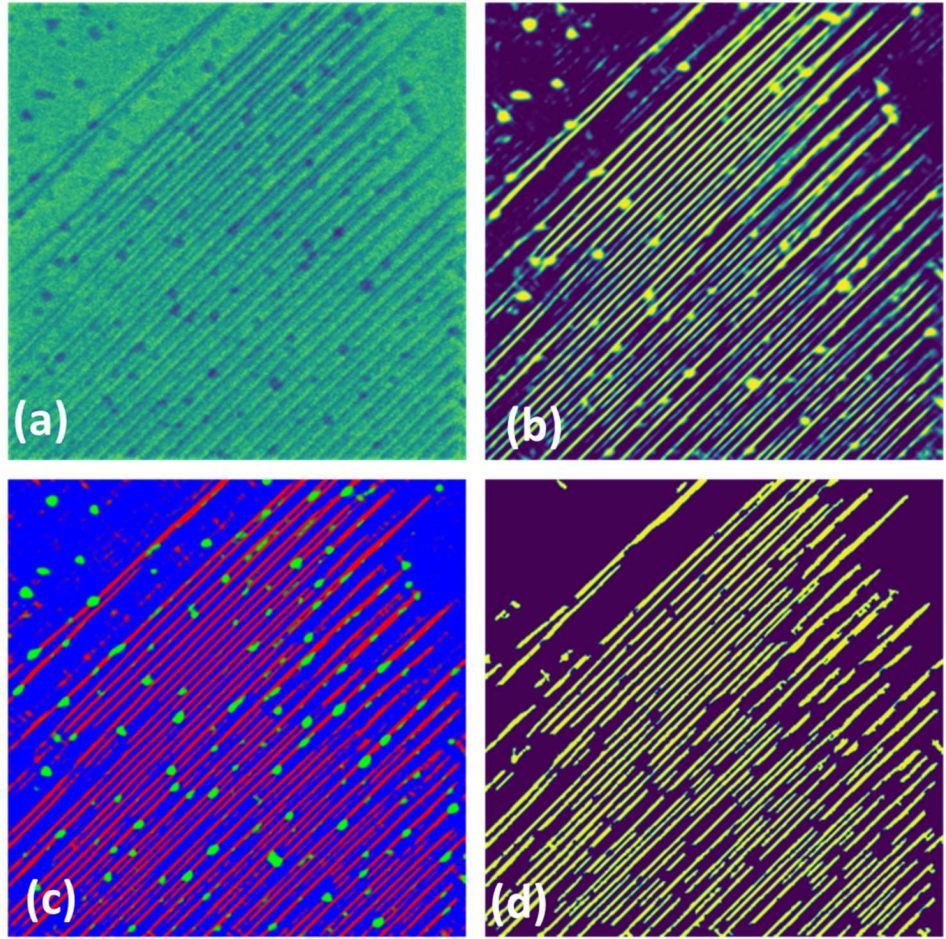

**Figure 2.** (a) Original (colored) BF-STEM image. (b) Result of semantic segmentation in 2 components, (c) semantic segmentation in 3 components, and (c) domain wall class image after semantic segmentation in 3 components.

      The DCNN based semantic segmentation of the STEM image data is illustrated in Figure 2. Here, the original BF-STEM image (artificially colored) is shown in Fig. 2 (a), clearly showing the domain walls and surface contaminants (in blue). The human-labeled training set was created for semantic segmentation of the STEM images, allowing for 2- and 3-component segmentation. In the former case, the DCNN was trained to recognize domains and domain walls. The results of this analysis are shown in Fig. 2 (b), where the contaminants are classified the same as the domain walls. To avoid this problem, the label set was extended to include both domains, domain walls, and contaminants as separate classes. In this case, the example of the semantically segmented image is shown in Fig. 2 (c) and clearly separates the relevant microstructural elements. The pure domain walls for this case are shown in Fig. 2 (d). Depending on the specific use case, either original images or decoded 2-class or 3-class images may be used for the subsequent analysis.

To determine the relevant domain geometries, we utilize the rVAE approach. In general, autoencoders are a class of neural networks where the original data set is compressed via a set of dense or convolutional layers into a small number of latent variables and subsequently deconvoluted into the original data set. Training of the autoencoders aims to minimize the reconstruction error. In this manner, latent variables create an optimal sparse representation of the initial data set, somewhat like a non-linear form of principal component analysis.

Variational autoencoders (VAEs) adopt the same principle, but in this case learn a generative model, i.e., they attempt to model the generating distribution of the dataset with the latent layer representing a Bayesian prior distribution.[36] Here, the stochastic encoder is used to approximate the true posterior of the generative model (the decoder).[37] Hence, in addition to minimizing a reconstruction loss between input and the generated output, one also minimizes the Kullback-Leibler divergence between the approximated and the true posterior distributions. Importantly, VAEs allow one to create low-dimensional continuous representations of the high-dimensional spaces in which the variation of latent variables in a certain direction allows disentangling the characteristic aspects of the object's behavior. Multiple examples of such analysis include writing styles for the MNIST (Modified National Institute of Standards and Technology) data base of handwritten digits and emotions for human face analysis. Additionally, VAEs generally subdivide the latent spaces between the classes present in the data, with arbitrarily drawn elements of the latent space potentially belonging to the extant class, extrapolation between, or being non-existent. These remarkable properties can be used for analysis of physical data, as performed here.

We adapted a version of the VAE optimized for the analysis of systems that contain features with general rotational symmetry,[38] *i.e.*, same object can appear at different orientations. In this case, three of the latent variables are rotation angle and offsets in *x*- and *y*- direction. The number of remaining latent variables is defined by the operator and here is typically set at 2. The rVAE's encoder had two convolutional layers with 128 filters ("kernels") of size (3, 3) in each layer activated by leaky-ReLU function with a negative slope of 0.1. The rVAE's decoder consisted of two fully-connected ("dense") layers with 128 neurons in each layer activated by *tanh()*. The encoder and decoder were trained simultaneously using Adam optimizer with the learning rate of 0.0001.

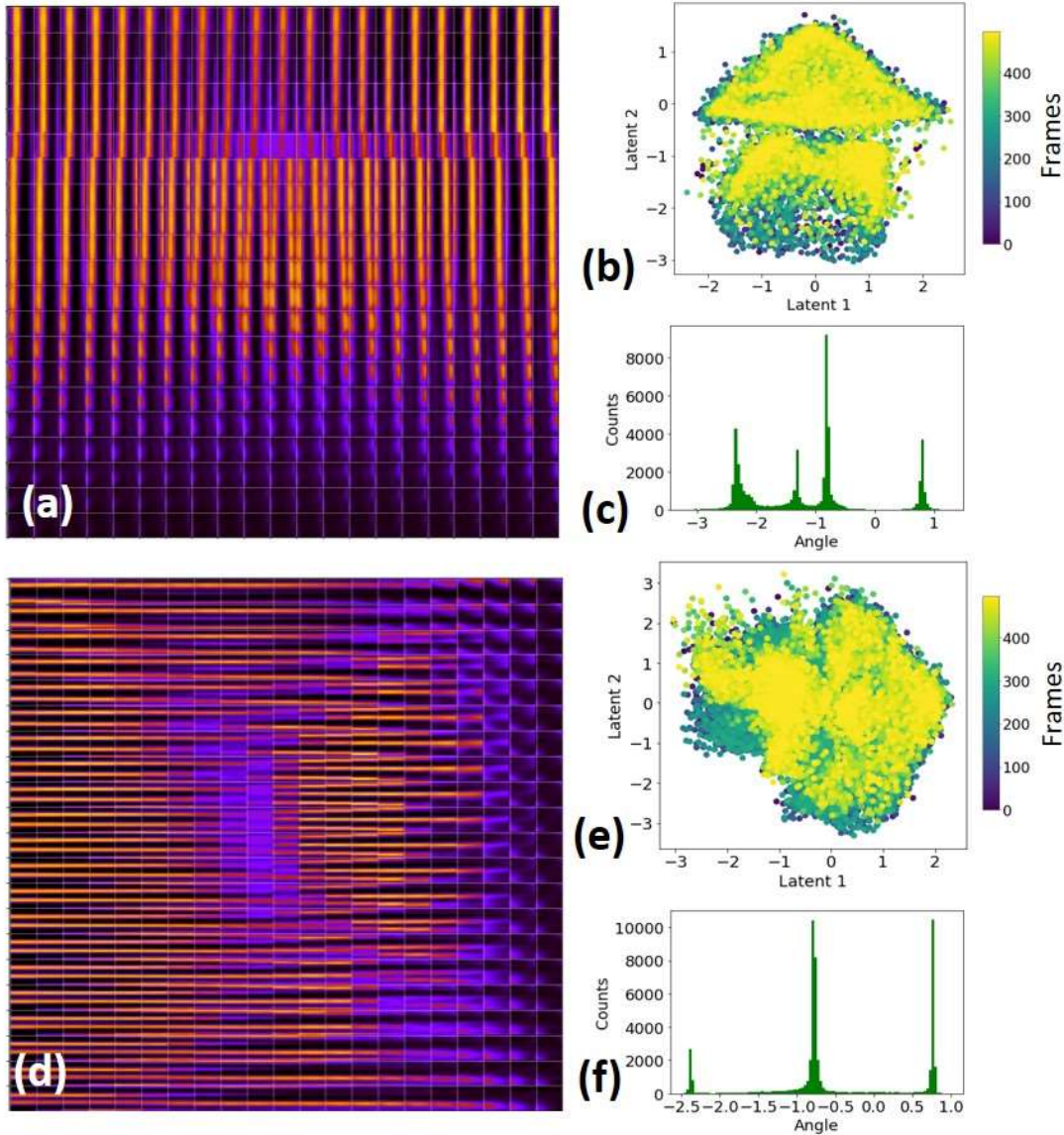

**Figure 3.** (a) Latent space of rVAE for window size $n = 8$. (b) Distribution of experimental points in latent space and (c) corresponding angle distribution (radians). (d) Latent space, (e) latent point, and (c) angle distributions for $n = 24$. Points in latent space in (b,e) are color coded according to time.

The rVAE analysis of the DCNN output of the data shown in Figures 1 and 2 are shown in Figure 3. Figure 3 (a,d) illustrates the behavior of the latent space for a small, $n = 8$, and intermediate, $n = 24$, size of sub-images (which we refer to as window size) randomly extracted from the semantically-segmented data. The diagram can be understood as follows: rVAE analysis encodes each sub-image of size $n$ x $n$ into five latent variables, namely the latent angle, $x$- and $y$-offsets, and two free latent variables, $L_1$ and $L_2$. Each sub-image defines a point in two-dimensional $L_1$, $L_2$ space. The complete set of sub-images defines the physically realizable region in latent space bounded by values ($L_{1min}$, $L_{1max}$) and ($L_{2min}$, $L_{2max}$). Note that in agreement with the rVAE design, the features in the latent space have identical orientation angle and thus, all variants of the ferroelectric walls can correspond to a single combination of latent variables, $L_1$ and $L_2$. However, the wall orientation and shift within the sub-image will be encoded in the latent angle and offsets, respectively.

The behavior of the latent space for a small window size is shown in Figure 3 (a). In this case, the latent angle and offsets encode the orientation of the domain wall and its position within the sub-image. At the same time, the $L_1$, $L_2$ variables have a very weak effect on the encoded domain shape, which at the first approximation corresponds to minor changes of the wall curvature. These behaviors can be further verified by analysis of the latent space distributions. For a small window size, the latent angle has a relatively broad distribution, which nonetheless contains peaks corresponding to the primary domain orientation variants within the image. The corresponding offset distributions are relatively narrow (not shown).

For the large window size, the dynamics captured by the rVAE become much richer. Here, the different parts of the latent space are occupied by dissimilar domain morphologies, e.g., double and triple domains. What is remarkable is that certain parts of the latent space, moving along selected directions recognizably, describe continuous changes in domain structure, e.g., an increase in the domain spacing or the emergence and growth of a new domain. At the same time, some parts of the latent space are populated by unphysical reconstructed structures that are readily recognizable by having an intermediate contrast. Note that since in this case the rVAE was applied to a semantically segmented image, we generally anticipate the contrast to be 0 or 1 with intermediate values confined only to small intermediate regions. The large regions with contrast adopting intermediate values do not correspond to physically realizable configurations. Also note that these regions will have very low density of data points from the original data set.

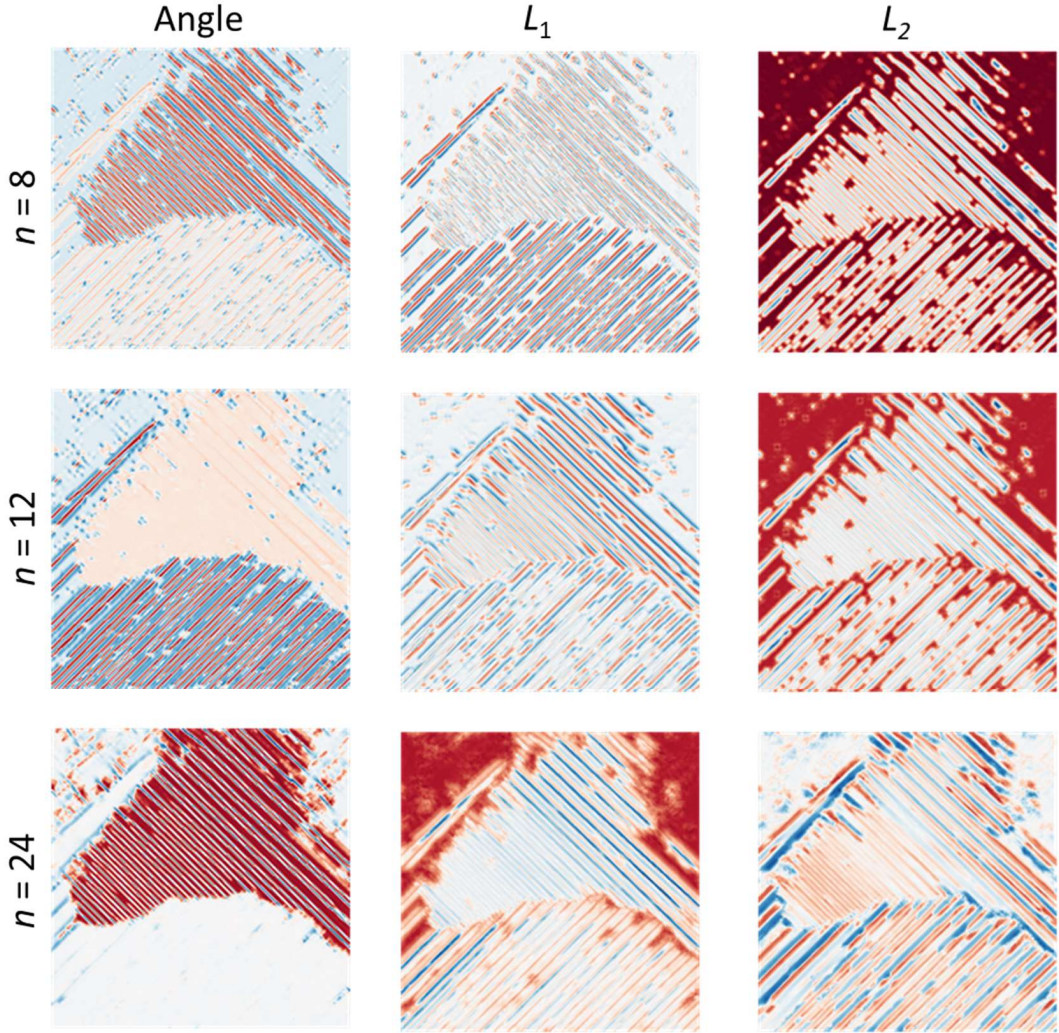

**Figure 4.** Latent angle and $L_1$, $L_2$ for three different window sizes.

Further insight into the behavior of the rVAEs can be obtained via analysis of the encoded images. Here, for each pixel of the image we form a sub-image centered on the image and use the rVAE trained on the full image stack to determine the latent angle, offsets, and latent variables corresponding to the image. Encoded images using the rVAE for three window sizes are shown in Figure 4. The original image shows a clear domain pattern in two variants and a number of surface contaminants of circular shape. Note that while these features are obvious to the human eye and are readily interpretable by a person familiar with the physics of ferroelectrics, such features are not familiar to the machine learning algorithm. The trained rVAE, however, clearly identifies the rotation angles when the features are recognized. The offset images contain the shadow of domain structures describing shifts of the domain walls within the sub-images. The latent images typically split with one of the latent variables encoding the presence of the domain walls and with the second encoding details of the domain structure.

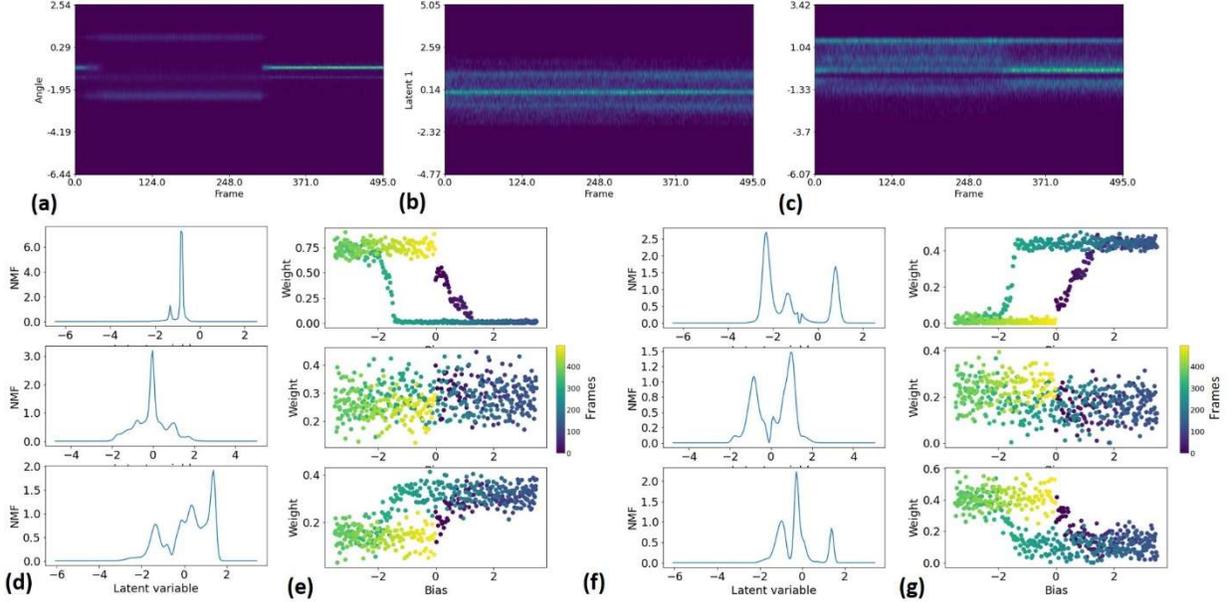

**Figure 5.** Evolution of latent variables during polarization switching and multivariate analysis for $n = 12$. (a) Latent angle, (b) $L_1$, and (c) $L_2$ as a function of frame index. (d) First NMF components of angle, $L_1$ and $L_2$ and (e) their bias dependence. (f) Second NMF component of angle, $L_1$, and $L_2$ and (g) their bias dependence. Note clearly visible transition hysteresis. Colors indicate frame.

The trained rVAE can be used to explore phase transition behavior via non-negative matrix factorization (NMF) analysis of the latent variables' statistics, similar to the GMM analysis reported earlier. Note that while for categorical GMM labels the data can in principle be represented via discrete variables (i.e., descriptor is the *M*-component vector describing the fraction of the *M*-th group of the GMM labels), the rVAE variables are continuous, necessitating such a representation. Here, each image during the experiment is encoded via rVAE, thereby giving rise to a set of latent variable values for each time step. The corresponding kerned density estimate is calculated, yielding the time-dependent probability density, e.g., for the $L_1$ variable, the time evolution is described via 2D function kernel density KDE($L_1$,t). The KDEs for the angle and two latent variables are shown in Fig. 5 (a-c). Note the clearly visible sharp changes in the time dependence of the KDEs, indicative of a first order bias-induced transitions.

This behavior can be further simplified via suitable dimensionality reduction, chosen here to be NMF, since NMF components are positively defined. In this case, the 2D KDE is represented as a sum of the product of weights representing the latent distributions and components representing their time dynamics. Here, we found that N = 2 allows adequate representation of the system dynamics. This behavior can further be plotted as a function of the applied bias (rather than time), as shown in Fig. 5 (e,g). Remarkably, the bias dependence of the angle and one of the latent components shows a clearly hysteretic character, whereas another latent component is almost bias independent.

The significance of this finding is that the combination of rVAE compression and subsequent matrix factorization is an unsupervised learning process with no *ad hoc* assumptions on the physics of the observed process. Despite this, the clear identification of relevant microstructural elements and identification of hysteretic behavior is possible. It is also important to note that while the exact rVAE behavior depends on the sampling window and training history, the overall behavior observed in Figure 5 is universal.

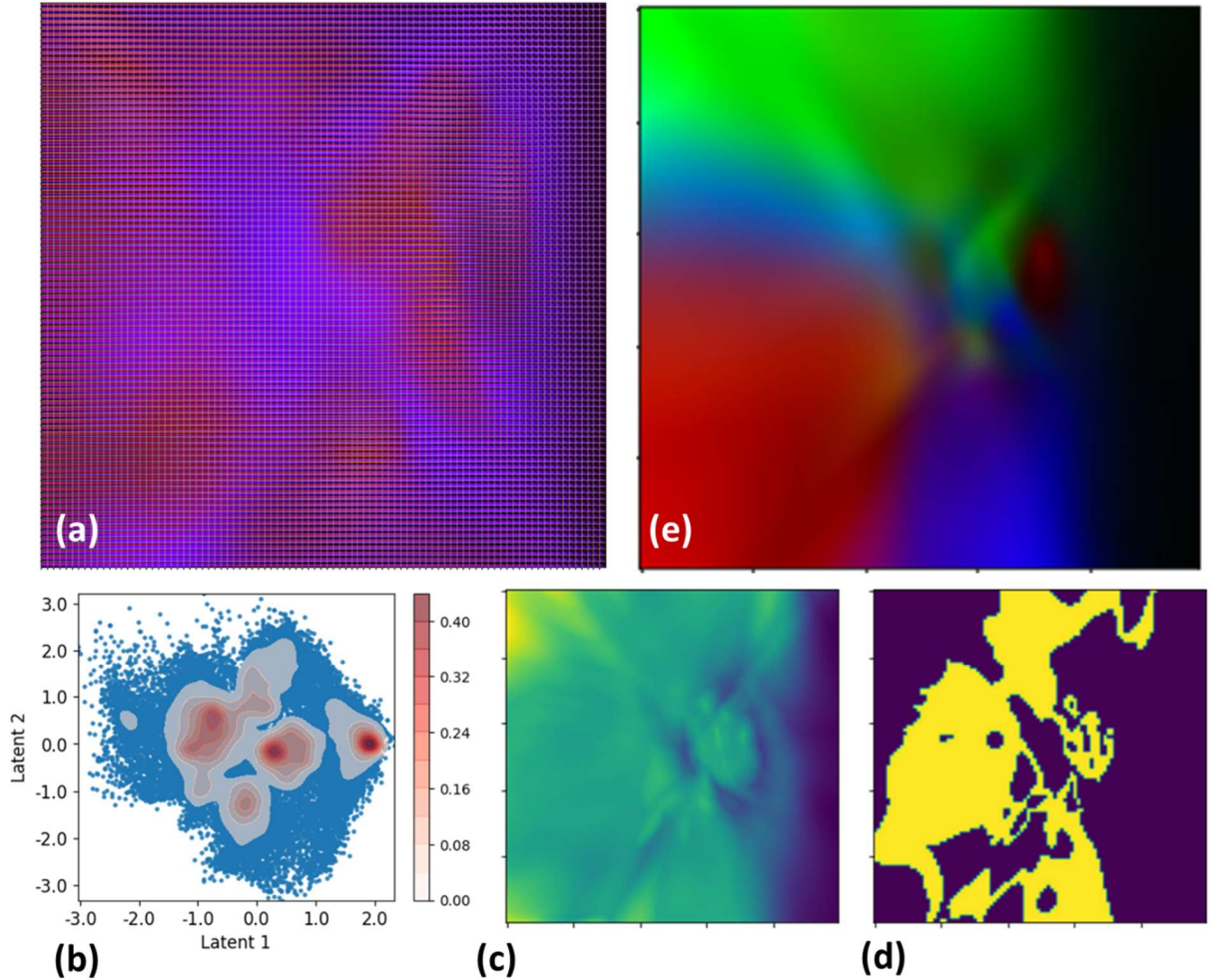

**Figure 6.** (a) Latent space of rVAE for $n = 24$ window size and dense sampling. Note clear large-scale contrast differentiating regions with physical and unphysical domain structures. (b) Corresponding distribution of observed points with superimposed probability density. (c) Maximum of reconstructed contrast and (d) regions where contrast maximum is within (0.95, 1.05). (e) RGB representation of latent domain morphologies.

Finally, we note that the structure of the latent space contains rich information on domain structures, morphologies, and transitions mechanisms. Close examination of the data in Fig. 3 (a,d) indicates the presence of well-defined domain structure elements in certain regions of the latent space, whereas other regions contain objects that are clearly unphysical, e.g., represent overlaps between multiple domain orientations. These behaviors can emerge since the original data sets form the certain distribution in the latent space and not all the points are populated. These manifolds represent interpolation between physically possible domain structures. Here, we expand upon these observations to map the regions with dissimilar physical behaviors in the latent space.

Figure 6 (a) shows the latent space for $n = 24$ window size for a dense grid of latent points. In this case the individual sub-images cannot be recognized; however, the overall image still has very discernible structure, which is due to the fact that the unphysical domain structures decoded from the latent space generally have weaker contrast and are observed as lighter regions. Note that the distribution of the original data in the latent space clearly forms discernible clusters, Fig. 6 (b). To obtain insight into this behavior, we map the maximal contrast of reconstructed domains as shown in Fig. 6 (c). The regions where this contrast is within (0.95, 1.05) is shown in Fig. 6 (d) and this illustrates the first approximation for the domains of physically realizable structures.

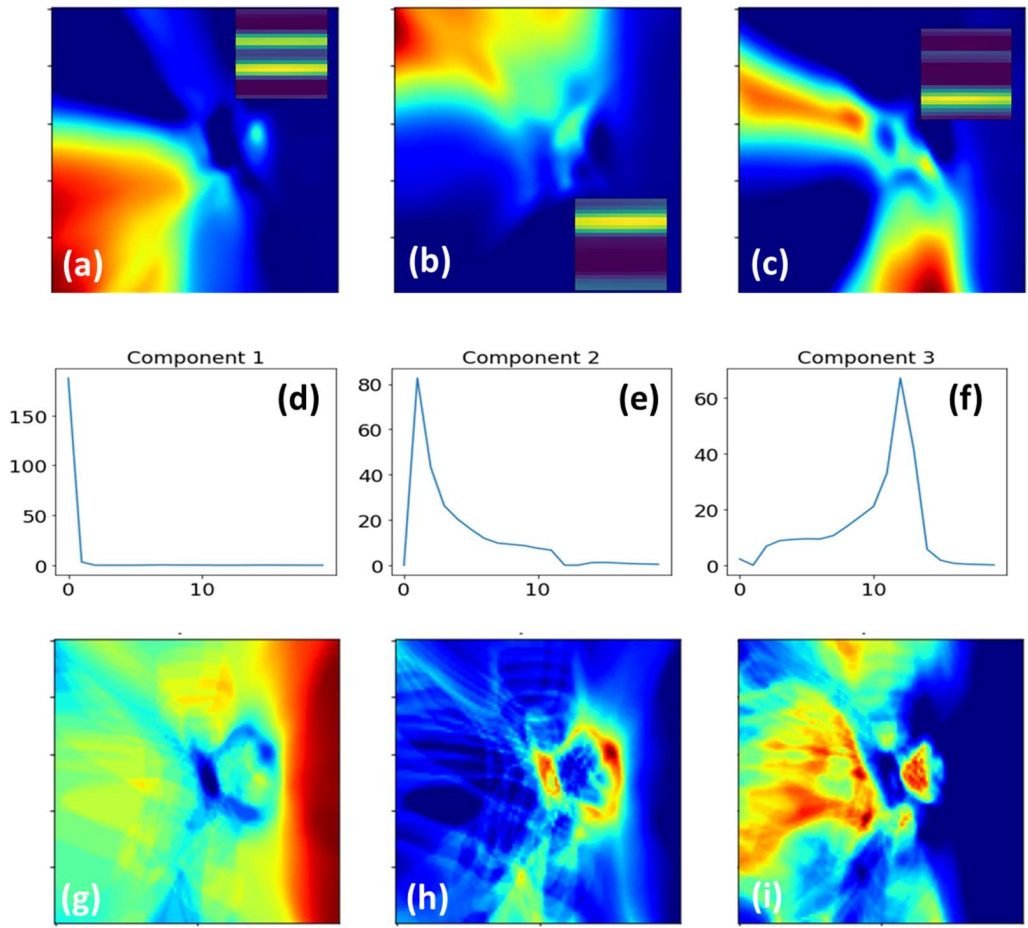

**Figure 7.** (a-c) NMF maps on reconstructed images. Shown are loading maps, with NMF components shown as insets. (d-i) NMF analysis of histograms. Shown are (d-f) components and (g-i) corresponding loading maps.

Next, we analyze the objects decoded from the latent space. In the first approach, we reconstructed the images from a uniform grid in the latent space and perform the classical NMF analysis. The results are shown in Fig. 7 (top row). Regions corresponding to double and single domain structures are clearly visible. Note that while somewhat similar information can be obtained from analysis of the original images, in this case we analyzed the low-dimensional manifold containing these objects and these data are represented as an RGB object in Fig. 6 (e).

Alternatively, the structure of the latent space can be explored via multivariate analysis of the histograms of the reconstructed images. For the physical image, the contrast distribution is beta-like and is dominated by the low (no domain wall) and high bins. At the same time, for the unphysical images, intermediate contrast scales are abundant. Therefore, examination of the NMF endmembers and the corresponding loading maps visualized in the latent space provide insight into regions of physical and unphysical behaviors. Ultimately, one would want to supplement the learned behaviors from model predictions of *a priori* physics knowledge on the range of possible states. After all, states that are in the data are obviously physical, but those that are predicted by the generative model may be either physical or unphysical, and preferably, a penalty could be added to the learning process to encourage the physical states over the unphysical ones. Moreover, if it is possible to quantify (roughly) the energy of individual states, one can find transition pathways in the latent space directly through pathfinding or reinforcement learning approaches. An alternative pathway would be to try to simulate the transitions directly in the generative model by 3D convolutions and image stacks as input to a 3D rVAE (i.e., working in ($x,y,t$)). Nonetheless, the key to making this possible is the realization of low-dimensional, learned descriptors that can effectively summarize energy-degenerate states within the same latent variables, as is physically relevant. In other words, rotational and shift invariance lets us to consider energy-degenerate states equivalently without additional complexity; in this case by encoding the symmetry constraints. However, this approach can be extended toward more complex physics.

To summarize, here we introduce a universal approach for analysis of dynamical electron microscopy data based on rotationally invariant latent embeddings and apply it for the exploration of bias-induced transformations in $BaTiO_3$ visualized via *in situ* STEM. The dynamic microscopy data contains information on phase transformation mechanisms in the form of domain structures and morphologies. Variational autoencoders with rotational invariance are used as a universal method to project the broad variety of domain patterns onto the low-dimensional latent space. The bias-dependence of latent variable distribution allows description of the transition mechanism, in this case yielding the hysteresis loop. Deep analysis of the latent space allows identification of regions with physical and unphysical domain configurations, with clearly identifiable evolutionary pathways within the physical regions.

The proposed approach is universal and can be applied to other dynamic phenomena ranging from thermal, bias, or chemically induced transformations in physical, chemical, electrochemical, and biological systems. Future developments can include exploration of dynamics and trajectories in the latent space for cases where individual physical units can be identified. Furthermore, we note that this approach is not limited to classical imaging and can be extended to multimodal and hyperspectral images, including microRaman, electron energy loss spectroscopy (EELS), and other imaging methods.

The dataset used to showcase the developed analysis was previously used in Physical Review Materials 4 (2020) 104403 and arXiv:2011.05842.

**Acknowledgements:** This effort (machine learning) is based upon work supported by the U.S. Department of Energy (DOE), Office of Science, Basic Energy Sciences (BES), Materials Sciences and Engineering Division (S.V.K.) and was performed and partially supported (M.Z.) at the Oak Ridge National Laboratory's Center for Nanophase Materials Sciences (CNMS), a U.S. Department of Energy, Office of Science User Facility. R. I. and V. T. acknowledge financial support from the Swiss National Science Foundation (SNSF) under award no. 200021_175711.